\documentclass[%
 reprint,%
 amssymb, amsmath,
 aps,prl,%
groupedaddress,%
]{revtex4-1}

\usepackage{docs}%
\usepackage{bm}%
\usepackage{graphicx,subfigure}
\usepackage{epstopdf}

\usepackage{color}

\begin{document}

\title{Temporal Loop Multiplexing: A resource efficient scheme for multiplexed photon-pair sources}%

\author{Robert J.A. Francis-Jones}%
\email{r.j.a.francis-jones@bath.ac.uk}

\author{Peter J. Mosley}

\affiliation{Centre for Photonics and Photonic Materials, Department of Physics, University of Bath, Bath, BA2 7AY, United Kingdom.}%

\date{\today}%

\begin{abstract}
Single photons are a vital resource for photonic quantum information processing. However, even state-of-the-art single photon sources based on photon-pair generation and heralding detection have only a low probability of delivering a single photon when one is requested. We analyse a scheme that uses a switched fibre delay loop to increase the delivery probability per time bin of single photons from heralded sources. We show that, for realistic experimental parameters, combining the output of up to 15 pulses can yield a performance improvement of a factor of 10. We consider the future performance of this scheme with likely component improvements.
\end{abstract}

\maketitle
Heralded single photon sources based on photon-pair generation in nonlinear media are at the forefront of developments in photonic quantum technologies~\cite{OBrien2007Optical-Quantum-Computing}. Photons are created in pairs, usually by a pulsed pump laser, allowing the delivery of a single photon in a well-defined time bin to be conditioned on the detection of its twin~\cite{Mosley2008Heralded-Generation-of-Ultrafast}. The simplicity of this approach is attractive, but, because pair generation is spontaneous, the generation probability must be kept low in order to limit contamination from higher-order photon-pair components~\cite{Broome2011Reducing-multi-photon-rates}. Combined with the effects of loss, the probability of delivering a heralded photon from any individual pump laser pulse is typically less than 1\%.

Realistic quantum information protocols require many single photon sources to be operated simultaneously~\cite{OBrien2007Optical-Quantum-Computing}. The probability of this occurring, $P$, for $N$ sources becomes vanishingly small if the individual probability of each source producing a photon $p \ll 1$ because $P = p^N$. One method of increasing $p$ is by multiplexing -- actively combining the output from several generation modes using delay, feed-forward, and a fast switch~\cite{Migdall2002Tailoring-Single-Photon,Shapiro2007On-Demand_Single,Ma2011Experimental_Generation_of,Collins2013Integrated_Spatial_Multiplexing}. Combining the output of pair generation in several separate sources -- spatial multiplexing -- brings significant performance benefits, but the resource overhead in nonlinear media and switches is high~\cite{Christ2012Limits-on-the-Deterministic,Adam2014Optimisation_of_Periodic}. 

Alternatively, one can combine the output from different temporal modes of the same source, bringing an overhead cost only in time and not in additional physical resources~\cite{Jeffery2004Towards-a-Periodic-Deterministic-Source,Mower2011Efficient_Generation_of,Glebov2013Deterministic_Generation_of,Mendoza2015Active_Temporal_Multiplexing,Schmiegelow2013Multiplexing-photons-with}. In this letter, we present a novel multiplexing scheme in the temporal domain based on a fibre loop that uses only a single optical switch and delay line. We numerically evaluate the performance of the scheme in the presence of imperfect detection, switch loss and attenuation in the delay line by extending the methods we presented in ~\cite{Francis-Jones2014Exploting_The_Limits} and consider the realistic limits to source performance.

\begin{figure}
	\includegraphics[width=8cm]{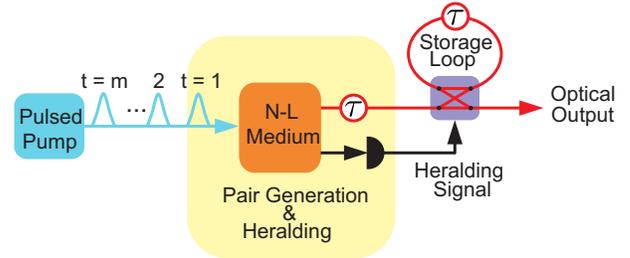}
	\caption{Schematic of temporal multiplexing scheme. See text for details.}
	\label{fig:TempMulit}
\end{figure}

Fig.~\ref{fig:TempMulit} shows the proposed setup. A non-linear medium ($\chi^{(2)}$ crystal or $\chi^{(3)}$ optical fibre) is pumped by a pulsed laser. The generated signal-idler pair is first split in wavelength, the signal photon is sent to a detector to herald the presence of the remaining idler photon. In this work, we consider a detector that has photon number resolving (PNR) capability, though the analysis could be equally well applied to a binary detector. The heralded idler photon is coupled into fibre and subjected to a fixed delay to give sufficient time for detection and feed-forward. Following the fixed delay a $2\times2$ optical switch is used to route the photon either to the output or into a fibre storage loop with a propagation delay of exactly one period of the the laser pulse train. Hence the probability amplitude of photons held in the storage loop will overlap with those of heralded idler photons from any subsequent laser pulse.

The multiplexing scheme is described as follows. A train of $m$ laser pulses, labelled from $t = 1$ to $t = m$ in order of arrival, enters the nonlinear photon-pair generation medium. If a pair is generated on the first pulse and the heralding detector fires, the switch is set to the crossed state so that the corresponding idler photon is routed into the the storage loop. On the next pulse if the heralding detector fires again the switch is once again set to the crossed state, simultaneously rejecting the photon from the previous pulse and storing the new photon in the loop. If there is no heralding event, then the switch remains closed and photon in the storage loop completes another pass through the switch. This process is repeated up to and including the $m$'th pulse and we refer to $m$ as the \textit{multiplexing depth}.

On the $m$'th pulse, if there is no successful heralding event then the switch routes the stored photon amplitude into the optical output. However, if there is a successful heralding event on the final pulse the switch allows the newly-generated photon straight through to the output. Hence the photon leaving the output following the $m$'th pulse may have passed through the switch and storage loop anywhere from $t = 1$ to $t = m$ times and therefore accrued a vastly different amount of loss. At low values of $t$ a photon must make more passes through the loop to be used after the $m$'th pulse and so its contribution to the overall probability of successfully delivering a single photon from the scheme is lower compared to photons generated on later pulses where $t \rightarrow m$. Furthermore, the loss of the switch and storage loop have a large effect on the probability of successfully delivering a single photon.

To evaluate the performance of the proposed system we follow a similar approach as outlined in~\cite{Francis-Jones2014Exploting_The_Limits}. We first assume that the photon-pair source has been engineered to generate signal and idler photons in only two spatio-temporal modes such that the modes are only correlated in photon number~\cite{Mosley2008Heralded-Generation-of-Ultrafast}. This allows us to describe the probability amplitudes of the state-vector,
\begin{equation}
	|\Psi\rangle = a_{0}|0_{s},0_{i}\rangle + a_{1}|1_{s},1_{i}\rangle + a_{2}|2_{s},2_{i}\rangle + \cdots,
	\label{eq:StateVector}
\end{equation}
with thermal statistics:
\begin{equation}
	|a_{n}|^{2} = p_{th}(n) = \frac{1}{(\bar{n} + 1)} \left(\frac{\bar{n}}{\bar{n} + 1}\right)^{n},
	\label{eq:ThermalStats}
\end{equation}
where $\bar{n}$ is the mean photon number per pulse.

A simple result can be obtained by only considering the first non-zero term in the state-vector expansion in Eq.~\ref{eq:StateVector}. The overall probability of successfully delivering a heralded single photon on the $m$'th pulse, $p(\text{success})$, is given by
\begin{equation}
	p(\text{success}) = 1 - \prod_{t=1}^{m}\left( 1 - p_{th}(1)\eta_{d}\eta_{L}^{t}\right),
	\label{eq:AnalyticRes}
\end{equation}
where $\eta_{d}$ and $\eta_{L}$ are the detector efficiency and lumped efficiency of one pass of the switch and storage loop. However, this expression is only valid at low values of $\bar{n}$. In order to describe the behaviour of the source more faithfully at higher mean photon numbers we have analysed the effects of higher photon-number terms from the state vector. To do so we applied the formalism developed in~\cite{Francis-Jones2014Exploting_The_Limits} in which we determined the effect of concatenated loss in spatially multiplexed single photon sources. We used this to calculate the reduced density matrix, $\hat{\rho_{i}}(n_{s},t)$, describing the heralded idler state given a heralding detection of $n_{s}$ photons on the $t$'th pulse and delayed until the $m$'th pulse. We modelled the effects of loss by applying a beam splitter transform in which the transmission coefficient is given by the lumped efficiency of $t$ passes through the switch and delay line. 

For each pulse $t$ there exists a set of normalised density matrices $\{\hat{\rho_{i}}(n_{s},t):t, n_{s} \in \mathbb{N}\}$ with $\text{Tr}\{\hat{\rho_{i}}(n_{s},t)\} = 1$, which are dependent on the number of photons $n_{s}$ detected by the heralding detector. By working with a PNR detector all detection results where $n_{s} \neq 1$ can be ignored. To extract the probability of successfully delivering a single photon from the $t$'th pulse we multiplied $\hat{\rho_{i}}(n_{s} = 1,t)$ by the probability of achieving a heralding detection result of $n_{s} = 1$ and calculated the overlap with a pure single photon Fock state:
\begin{equation}
	p(\text{success},t) = p(n_{s} = 1)\frac{\langle 1|\hat{\rho_{i}}(n_{s} = 1,t)|1\rangle}{\text{Tr}\{\hat{\rho_{i}}(n_{s} = 1,t)\}}.
\end{equation}
The overall probability of successfully delivering a single photon from a pulse train divided into different temporal bins, each containing $m$ pulses is found through the product
\begin{equation}
	p(\text{success},m) = 1 - \prod_{t = 1}^{m}\left(1 - p(\text{success},t)\right).
\end{equation}
The contribution to noise, defined as a successful heralding detection followed by the delivery of an idler state containing more than one photon, was found through,
\begin{equation}
	p(\text{noise},m) = 1 - \prod_{t = 1}^{m}\left( 1 - \sum_{n = 2}^{\infty} \frac{\langle n|\hat{\rho_{i}}(n_{s} = 1,t)|n\rangle}{\text{Tr}\{\hat{\rho_{i}}(n_{s} = 1,t)\}}\right),
\end{equation}
When $\hat{\rho_{i}}$ is properly normalized such that $\text{Tr}\{\hat{\rho_{i}}\} = 1$, this reduces to:
\begin{equation}
	p(\text{noise},m) = 1 - \prod_{t = 1}^{m}\left( \sum_{n = 0}^{1}\langle n|\hat{\rho_{i}}(n_{s} = 1,t)|n\rangle\right),
\end{equation}
it is nevertheless important to note that the accuracy of both $p(\text{success},m)$ and $p(\text{noise},m)$ will depend on the value of $n$ at which the calculation is truncated due to the effect this has on the normalization of $\hat{\rho}_i$. We define a corresponding signal-to-noise ratio,
\begin{equation}
	\text{SNR} = \frac{p(\text{success},m)}{p(\text{noise},m)},
\end{equation} 
that allows us to make direct comparisons of systems with disparate characteristics. Using these techniques, we have simulated numerically the effects of higher-order photon-pair components on a temporally-multiplexed source to enable the optimization of real-world systems in the presence of loss.

\begin{figure}
	\begin{center}
		\subfigure[]{
			\includegraphics[width=0.22\textwidth]{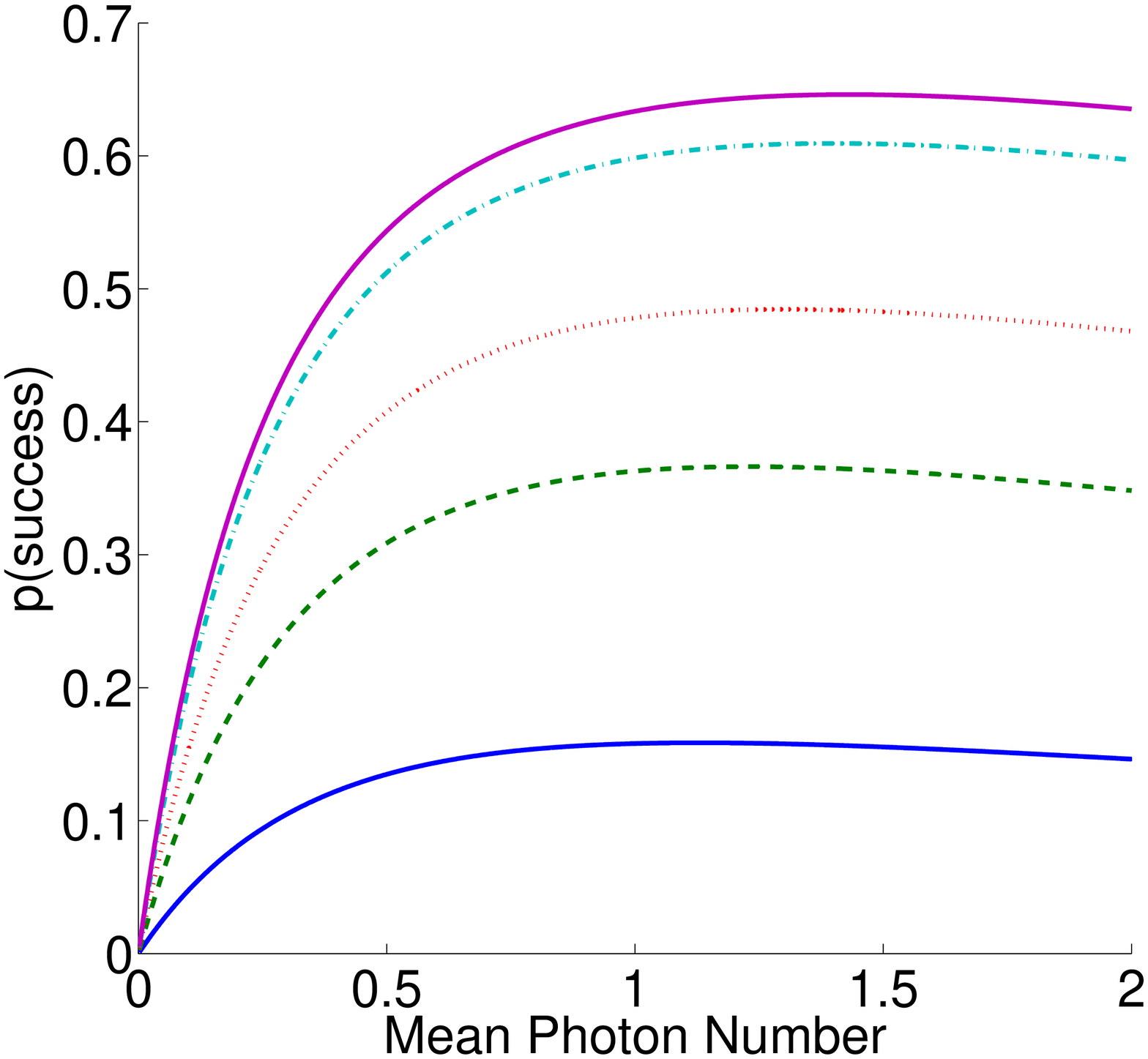}
			\label{fig:psuccess_nbar}
			}
		\hspace*{\fill}
		\subfigure[]{
			\includegraphics[width=0.22\textwidth]{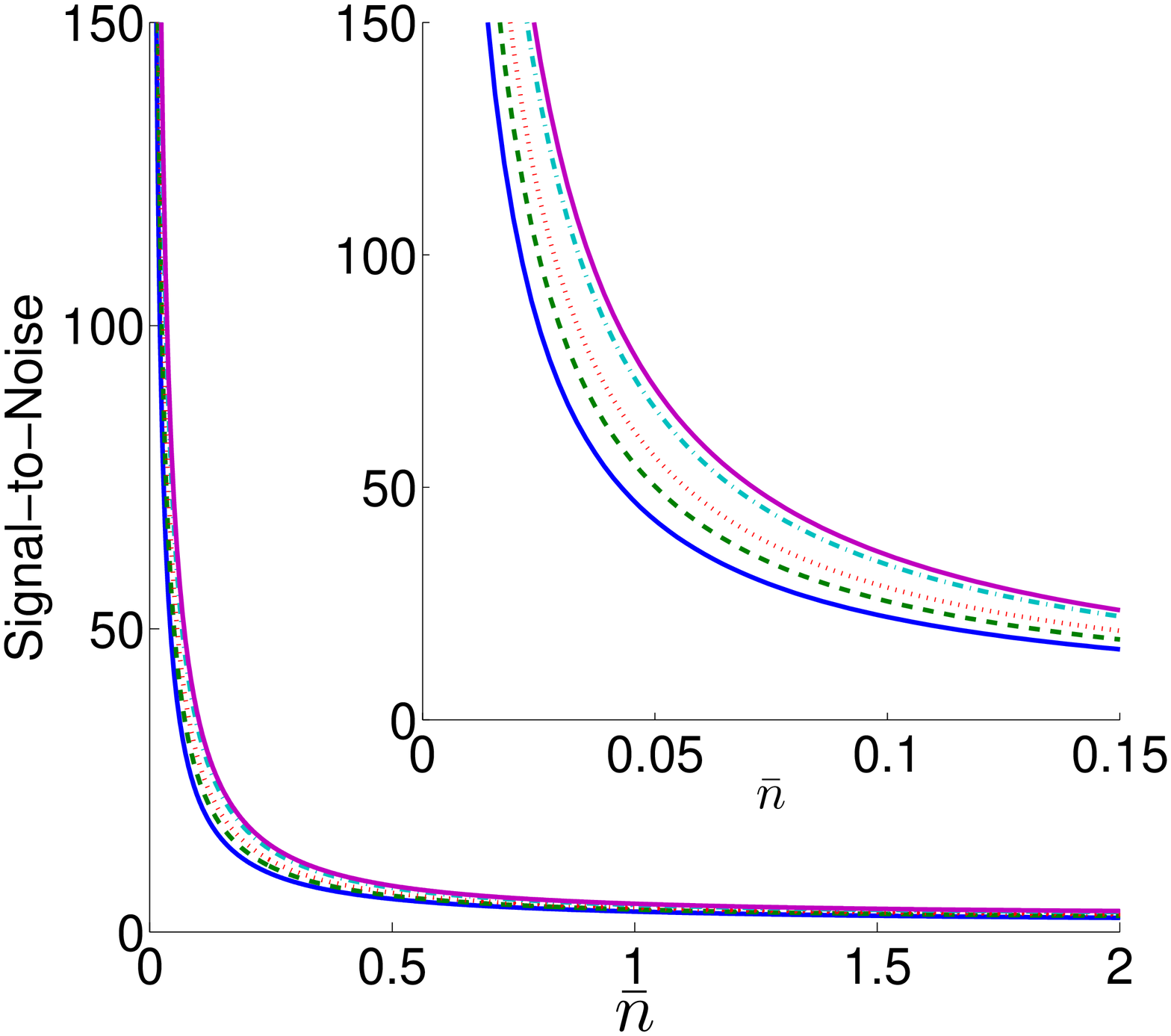}
			\label{fig:SNR_nbar}
			}
			
		\subfigure[]{
			\includegraphics[width=0.22\textwidth]{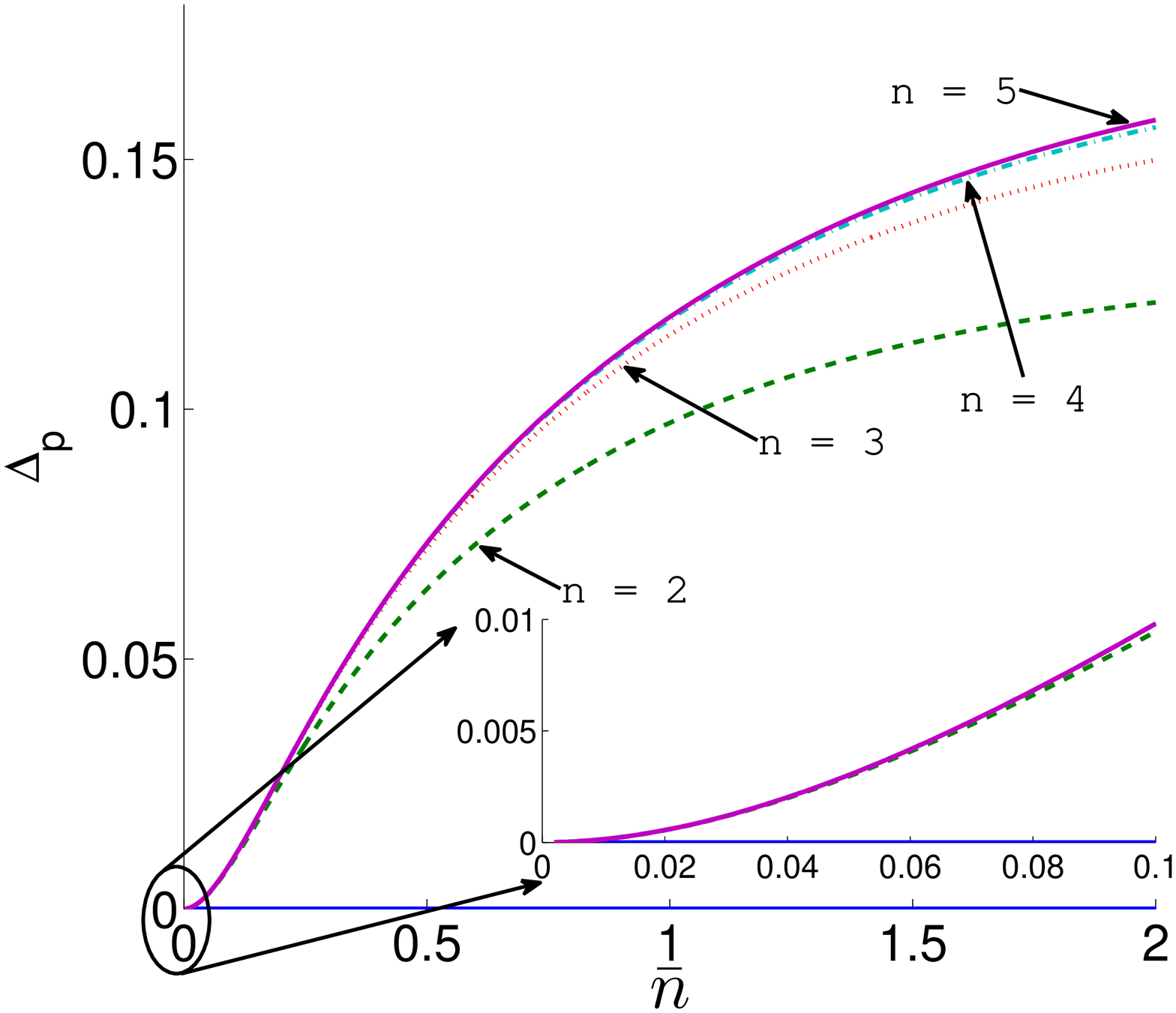}
			\label{fig:Difference}
			}
		\hspace*{\fill}
		\subfigure[]{
			\includegraphics[width=0.22\textwidth]{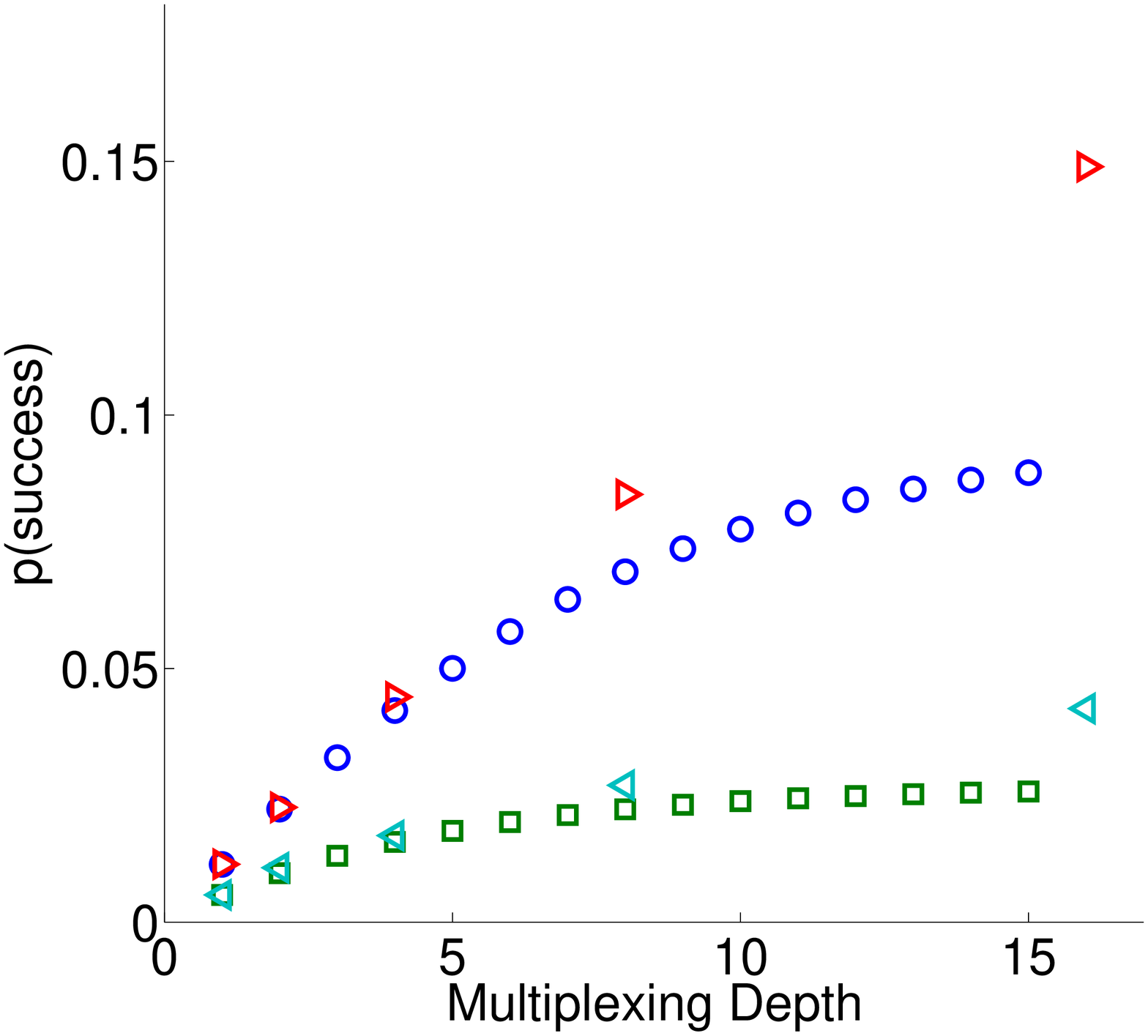}
			\label{fig:psuccess_mdepth}
			}
		\caption{a.) Variation in $p(\text{success})$ with increasing $\bar{n}$ for multiplexing depths of $m = 1$ (blue), 3 (green), 5 (red) and 15 (purple) pulses. b.) Corresponding signal-to-noise ratio. c.) Difference ($\Delta_{p}$) between $p(\text{success})$ calculated using Eq~\ref{eq:AnalyticRes} and the full numerical calculation with up to $5$ photon number components. Need to label the lines somehow d.) The increase in $p(\text{success})$  with increasing multiplexing depth at a fixed $\bar{n}$ (green squares) and fixed $\text{SNR} = 100$ (blue circles). Triangles show the comparison with a spatial multiplexing scheme for fixed $\bar{n}$ and fixed SNR in teal and red respectively. Simulations carried out with a detector efficiency $\eta_{d} = 0.7$ and lumped loop efficiency $\eta_{L} = 0.8$.}
		\label{fig:FourPanel_nbar}
	\end{center}
\end{figure}

Figure~\ref{fig:FourPanel_nbar} shows the performance of the proposed multiplexing scheme with $\eta_{d} = 0.7$ and $\eta_{L} = 0.8$ for different multiplexing depths $m$. As the multiplexing depth is increased the resulting probability of successfully delivering a single photon from the output increases. The corresponding SNR is shown in Fig.~\ref{fig:SNR_nbar}. To maintain a high SNR the mean photon number is constrained to be low in order to reduce the contribution from multi-photon components in the heralded density matrix. By multiplexing, the overall probability of successfully delivering a single photon at the output is increased at fixed SNR, as seen in Fig.~\ref{fig:psuccess_mdepth}.

It is insightful to find the limits of validity to the simple analytic result of eq.~\ref{eq:AnalyticRes} and thence to set the point at which our numerical calculations can safely be truncated. The difference between the analytic result and the numerical density matrix method truncated at different numbers of photon pairs is plotted in Fig.~\ref{fig:Difference}. We can see that for larger mean photon numbers it is essential to include higher-order terms in the density matrix, however, the benefit in using any amplitudes above 5 photon pairs quickly diminishes. As a result, all subsequent calculations were made by truncating the state vector at $n = 5$.


Figure~\ref{fig:psuccess_mdepth} shows the increase in $p(\text{success})$ with multiplexing depth $m$. Results from our previous simulations of a spatial multiplexing scheme are shown for comparison~\cite{Francis-Jones2014Exploting_The_Limits}. Our numerical approach allows a direct comparison between multiplexed devices at a fixed SNR of 100.  It can be seen that the temporal multiplexing loop yields an increase in $p(\text{success)}$ commensurate with spatial multiplexing schemes of approximately the same depth, but with a large reduction in the resource overhead required for implementation. In the case of temporal multiplexing the exponentially diminishing gain in $p(\text{success})$ with increasing passes through the storage loop leads to the improvement saturating at larger values of $\bar{n}$ as shown in Fig.~\ref{fig:psuccess_mdepth}. For small values of $\bar{n} < 0.01$, as the multiplexing depth increases, $p(\text{success})$ tends to a constant value given by
\begin{equation}
	\lim_{m\rightarrow\infty}\{p(\text{success},m)\} = \left(\frac{\eta_{L}}{1-\eta_{L}}\right)p(\text{success},1).
\end{equation}
This is in contrast to spatially-multiplexed sources, in which the probability of success continues to increase with multiplexing depth over the range studied here; nevertheless, as shown in~\cite{Francis-Jones2014Exploting_The_Limits}, even in the spatial case $p(\text{success})$ will begin to decrease at a value dependent on the concatenated switch loss.

At a fixed SNR of 100, a storage loop and switch with total loss $\eta_{L} = 0.8$ enables a factor of 6 increase in performance at a multiplexing depth of $m = 8$ pulses. This requires only one photon-pair source, one detector, and one switch. To achieve a similar increase in performance from a spatially-multiplexed scheme would require vastly more resources: 8 sources, 8 detectors, and 7 switches and delay lines. Not only is building such a network complicated and expensive, but the sources and delays must be carefully matched to ensure that the photons output are in identical pure states. The resource scaling for the schemes discussed is outlined in Table~\ref{table:Comparison}.
\begin{table*}
	\caption{Multiplexing scheme performance comparison. Improvement calculated for multiplexing depth $m = 8$, relative to single source with a heralding detector of efficiency $\eta_{d} = 0.7$ and switch efficiency $\eta_{L} = 0.8$.}
	\centering
	\begin{tabular}{c|c|c|c|c|c|c|c}
	
	\hline\hline
		Scheme & Sources & Heralding Detectors & Switches & Rep. Rate & SNR & $\bar{n}$ &  Improvement\\
	\hline
	Temporal & $1$ & $1$ & $1$ & $R_{p}/d$ & $100$ & - & $6.06$\\
	Spatial & $2^{d}$ & $2^{d}$ & $2^{d} - 1$ & $R_{p}$ & $100$ & - & $7.40$\\
	\hline
	Temporal & $1$ & $1$ & $1$ & $R_{p}/d$ & - & 0.01 & $3.32$\\
	Spatial & $2^{d}$ & $2^{d}$ & $2^{d} - 1$ & $R_{p}$ & - & 0.01 & $4.03$\\
	\end{tabular}
	\label{table:Comparison}
\end{table*} 

\begin{figure}
	\begin{center}
		\includegraphics[width=0.45\textwidth]{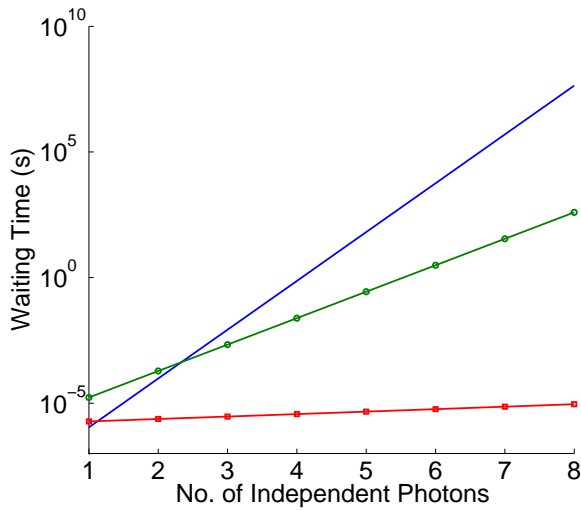}
	\end{center}
	\caption{The waiting time to deliver $N$ single photons from $N$ independent sources at fixed $\text{SNR} = 100$. Blue (solid) - Shows the waiting time for a standard single-photon source pumped at $80$MHz. Green (circles) - A multiplexing scheme with detector efficiency $\eta_{d} = 0.7$, switch efficiency $\eta_{L} = 0.8$ and a multiplexing depth of 15. Red (squares) - A future multiplexing scheme with detector efficiency $\eta_{d} = 0.98$, switch efficiency $\eta_{L} = 0.95$ and a multiplexing depth of 15. For large numbers of independent photons, these temporal multiplexing schemes exhibit a distinct advantage over traditional single photon sources.} 
	\label{fig:WaitingTimes}
\end{figure}

We note that after temporal multiplexing has taken place the number of time bins in which a photon can be delivered has been reduced by a factor of $m$. Therefore, in terms of number of heralded single photons delivered in one mode per second, this scheme may be outperformed by non-switched photon-pair sources pumped by a high-repetition-rate laser ~\cite{Morris2014Photon-Pair_Generation}. Nevertheless, if our goal is to deliver several single photons from independent sources simultaneously the scheme will yield huge improvements over current source architectures. This is summarized in terms of the waiting time to deliver $N$ single photons from $N$ independent sources shown in Fig.~\ref{fig:WaitingTimes}.

It is clear that to construct the highest performance single photon sources the loss of all the components between the point of generation and the output must be minimised. Switch loss is critical in this endeavour. Notwithstanding this, our analysis shows that, in the light of the inevitable imperfections in currently-available components, meaningful gains in source performance can be readily achieved by implementing this scheme.

\begin{figure}
	\begin{center}
		\subfigure[]{
			\includegraphics[width=0.22\textwidth]{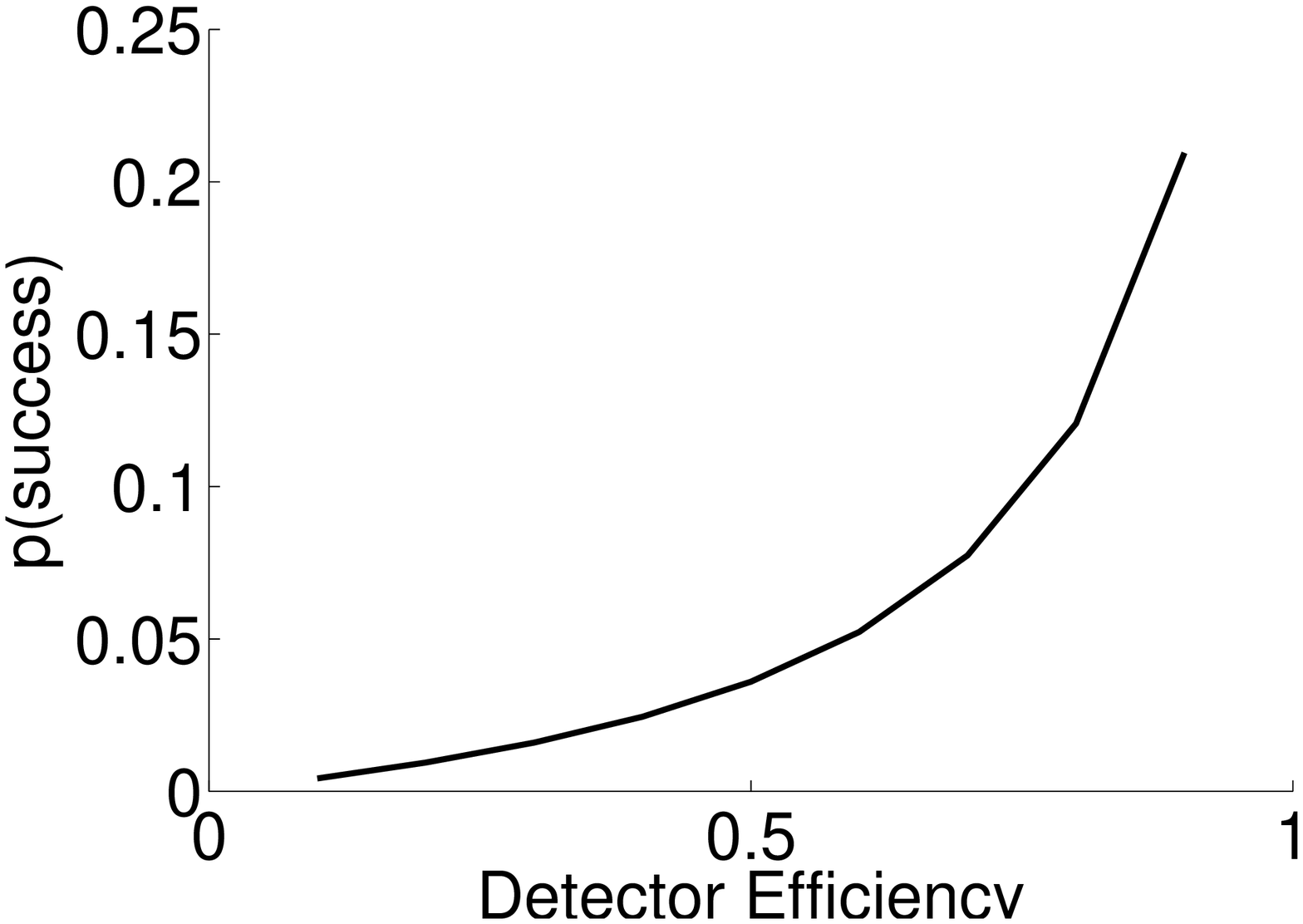}
			\label{fig:psuccess_etad}
			}
		\hspace*{\fill}
		\subfigure[]{
			\includegraphics[width=0.22\textwidth]{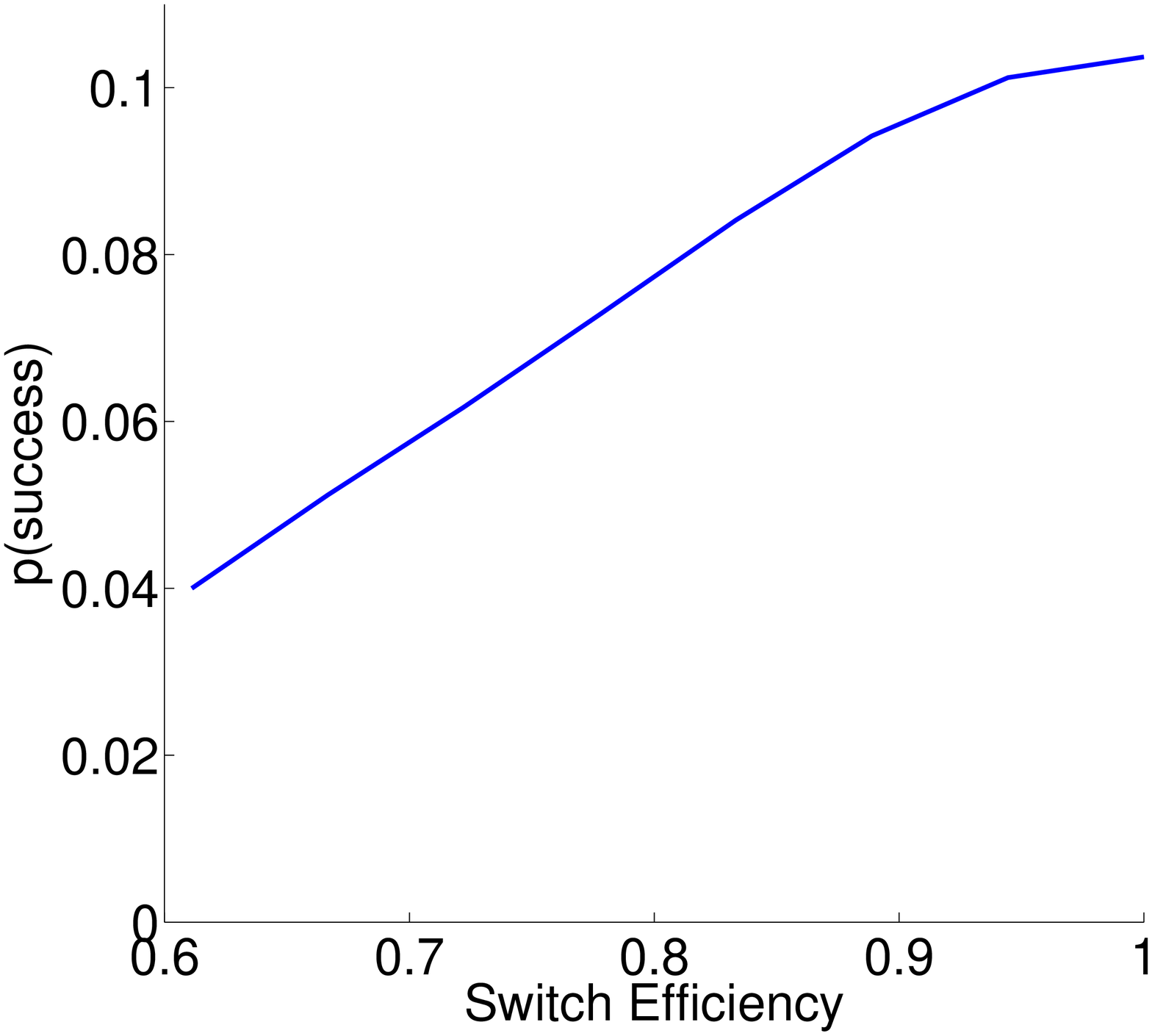}
			\label{fig:psuccess_nL}
			}
			
		\subfigure[]{
			\includegraphics[width=0.22\textwidth]{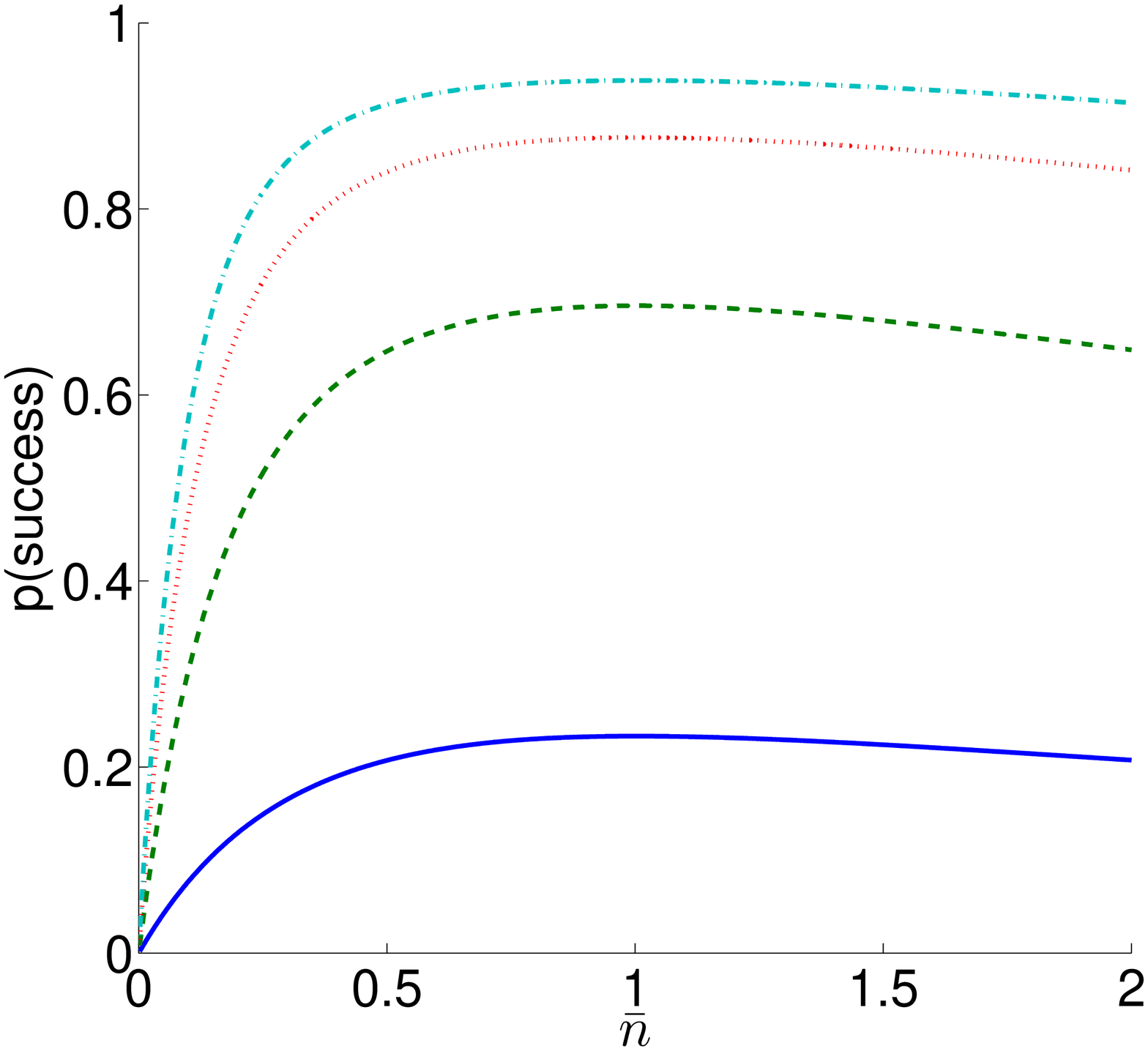}
			\label{fig:psuccess_nbar_higheff}
			}
		\hspace*{\fill}
		\subfigure[]{
			\includegraphics[width=0.22\textwidth]{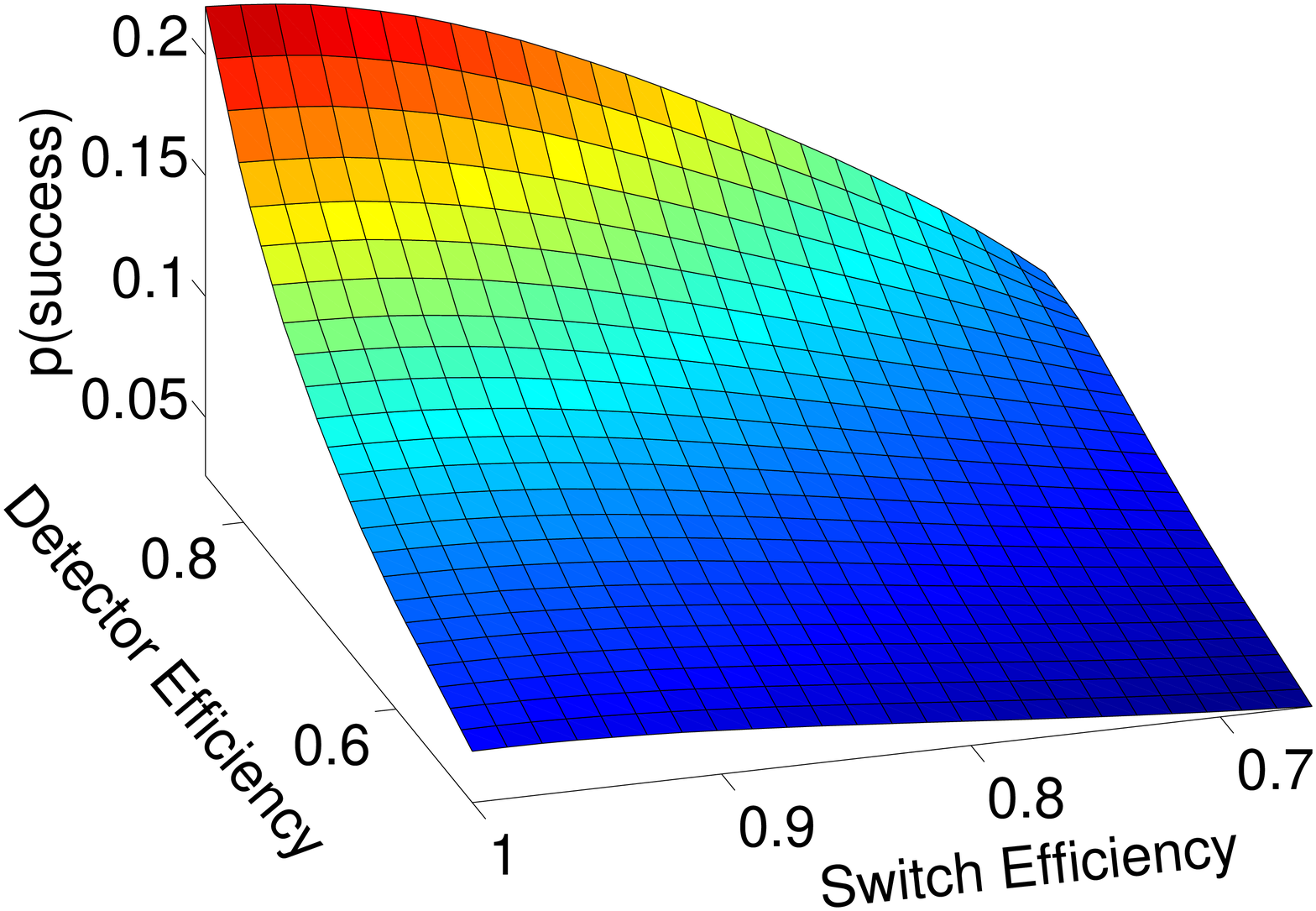}
			\label{fig:interpolated_efficiency_surface}
			}
		\caption{a.) Variation of $p(\text{success})$ with detector efficiency, $\eta_{d}$, for a fixed switch efficiency ($\eta_{L} = 0.8$) at fixed $\text{SNR} = 100$. b.) Variation of $p(\text{success})$ with switch efficiency, $\eta_{s}$, for a fixed detector efficiency ($\eta{_d} = 0.7$) at fixed $\text{SNR} = 100$. All points correspond to multiplexing depth of 10. c.) Realistic future device with $\eta_{d} = 0.98$ and $\eta_{L} = 0.95$, multiplexing depth 1 (solid), 5  (dashed), 10 (dotted) and 15 (dot-dash) which approaches deterministic operation. d.) Probability of success surface plot showing the overall effect of component efficiency for a multiplexing depth of 10 at a fixed $\text{SNR} = 100$.}
		\label{fig:TwoPanelEfficiencies}
	\end{center}
\end{figure}

Our method of analysis at a fixed SNR enables the effects of detector inefficiency and switch loss to be straightforwardly compared. Fig.~\ref{fig:TwoPanelEfficiencies} shows the behaviour of $p(\text{success})$ with varying detector and switch efficiencies at a fixed $\text{SNR} = 100$ at a fixed multiplexing depth of 10.

We see in Fig.~\ref{fig:psuccess_etad} that at low values of $\eta_{d}$ the detector has limited ability to discriminate between pulses on which a single pair or multi-pair generation event has occurred. The source must then be operated at very low mean photon number in order to maintain the signal to noise, severely limiting the success probability. As the detector efficiency increases, the probability of success grows rapidly due to the larger value of mean photon number that can be accessed; as $\eta_d \rightarrow 1$ the source can be operated where the one-pair component of the thermal distribution of photon number peaks. In this regime we are only limited by loss in the switch and the attenuation of the fibre in the storage loop. In the case of Fig.~\ref{fig:psuccess_nL}, provided the switch has an efficiency greater than $0.5$ then multiplexing is beneficial, however if the switch efficiency is not much greater than this limit, multiplexing over a large number of pulses will be not be worthwhile as the switch loss will throttle any improvement. Conversely, if the efficiency of the switch is high then the scheme will be limited by the efficiency of the heralding detector; the source should be pumped at a mean photon number that is low enough to limit the contribution of multi-photon events to the output, but multiplexed over a large number of pulses to raise the overall probability of success. Fig.~\ref{fig:psuccess_nbar_higheff} shows the potential for a future realistic device, with optimised detector and switch efficiencies to deliver a near deterministic single photon source using this novel architecture. 

In conclusion, we have shown that a temporal loop multiplexing scheme is an attractive prospect for increasing the performance of heralded single photon sources whilst minimising the number of components and the complexity of the complete device. Our analysis shows that to realise the benefits of multiplexing, the pump must be correctly tailored to the loss of the device. Therefore, numerical simulations such as those carried out above are key to extracting the maximum performance of these systems. By working in an all-fibre architecture loss can be minimised while ensuring straightforward integration.

\section*{Acknowledgements}
We acknowledge support from the UK Engineering and Physical Sciences Research Council under grant EP/K022407/1.

During the preparation of this manuscript we became aware of related work by Rohde et al ~\cite{Rohde2015Multiplexed_Single-Photon}.

\bibliography{BibDeskBibliography_NonEdit}

\end{document}